# On the origin of permeative flows in cholesteric liquid crystals

Weichao Zheng,[1,*]

[1]*Department of Chemistry, Physical and Theoretical Chemistry Laboratory, University of Oxford, Oxford OX1 3QZ, United Kingdom*

Permeative flows, known for the explanation of the anomalous viscosity ($10^5$ Poise) in cholesterics at low shear rates, are still under debate due to the difficulty of experiments. Here we use the Surface Force Balance, in which uniform domains with regular circular defects are formed, to probe the forces generated by compression in the direction of the helical axis. At the quasi-static speed of the surface approach, the measured forces are shown to be elastic (not dissipative), arising from the twist elastic deformation when the planar anchoring at the walls is strong. A mechanism involving frictional surface torque under strong planar surface anchoring will be proposed. The results indicate that the strong resistance to flow observed, previously interpreted as an enormous apparent *viscosity*, may in fact originate from the intrinsic non-linear increase of *elasticity* when the molecules are rotated away from equilibrium. The system is found to store energy (the force is reversible), without dissipation, as long as the applied stress is below the threshold for nucleating new defects. Our study underpins the importance of boundary conditions that may dramatically change the rheology of other viscoelastic materials and sheds light on the rational design of strain-stiffening materials, nanomotors, and artificial muscles involving helical architectures.

## 1   Introduction

Viscoelastic materials can both dissipate and store energy under stress, distinct from normal liquids and solids. Like the wave-particle duality of light, viscoelasticity has also incurred many debates, particularly in biology that is made of complicated physicochemical environments. For example, the relaxation time of membrane tension has been reported with a difference of six orders [1]. A recent study [2] reported that condensed chromatin with helical coiling is a solid supporting liquid protein condensates in vivo, which is different from a previous result of liquid chromatin in vitro [3]. Similarly, in a much simpler system, i.e., liquid crystals, the viscosity of cholesterics increases by five orders in the capillary [4], called permeative flows [5].

In chiral nematics, i.e., cholesterics, the molecules rotate periodically, forming long-range helical structures similar to double-helix DNA [6]. It is this molecular arrangement that gives rise to a long-standing puzzle: at low shear rates, the viscosity in cholesterics increases by $10^5$ times compared to that in nematics [4]. Many theoreticians [5,7-12] have agreed that the pressure along the helical axis, with an anchoring effect at the wall, induces the flow of molecules through the fixed helix, i.e., permeative flows, resulting in high apparent viscosities. Although some experiments [4,13-21] have been conducted to measure viscosity or probe permeative flows, by varying parameters such as the geometry of confinement [4,13,16,19,20], the direction of the helical axis [14,15,17], or the pitch length [18,21], these experiments all focused on macroscopic rheological behaviors without looking into detailed structures, leading to inconsistent results. In particular, due to the experimental difficulty [10-12,19,22,23], details on how boundary conditions, defects, and helical molecule dispersions interact with imposed forces were not explicitly characterized.

Here, we utilized the Surface Force Balance (SFB, Fig. 1c), in which the optical and mechanical properties of cholesterics can be explored simultaneously [24], to subtly measure forces due to tiny changes in the rotation rate of cholesterics under confinement. With determined boundary conditions, namely the anchoring strength and direction, the equilibrium forces during the surface approach were investigated using uniform cholesterics with regular circular defects, simultaneously taking into account the dynamic helical arrangement of molecules. With strong surface anchoring, cholesterics deformed elastically, which is reversible, behaving as a solid. When the measured elastic forces were assumed to be equivalent viscous forces, anomalous viscosity of cholesterics was obtained, which we believe is the origin of permeative flows, resulting from the strain-stiffening nature of the helical structures.

## 2  Mechanical winding of cholesterics by surface torque

Permeative flows are typically described as flows along the helical axis of cholesterics confined in a capillary (Fig. 1a) or between two plates (Fig. 1b). Here we used a third type of geometry, i.e., crossed cylinders in the SFB, to compress cholesteric layers also along the helical axis (Fig. 1c). With strong surface anchoring, cholesteric layers (layer thickness equal to half-pitch $P/2 = 122$ nm) deformed with more than 70% strain from 1140 nm to 300 nm during the approach of the surfaces, generating forces of more than 30 mN/m, before all layers were squeezed out from the contact position (Fig. 1d). The force profile is well fitted by Equation S4 with a strong anchoring (Fig. 1d), which manifests the mechanical winding of cholesteric layers without changing the total twist angle or being squeezed out layer-by-layer. During the jumping event, the slope is parallel with the spring constant $k = 179$ N/m (Fig. 1d), indicating the common mechanical instability [24] when the spring is softer than the sample.

This mechanical deformation of cholesterics is consistent with the constrained winding in our previous study [24]. Below the jumping threshold for nucleating new defects, the deformation can sustain for more than one hour without dissipation when the motor stopped [24]. However, the sustained forces could also result from a large viscosity of permeative flows. Here, unambiguously, we further confirm that the encountered force was reversible during the approach and retraction with the same motor speed (Fig. 1, g and i, and Fig. S1). Initially, the surface approached from 1450 nm to 700 nm staying for a short time, then retracted to 1500 nm, while the maximum applied strain was around 50%, less than the jumping threshold of 70% strain. The experiment was performed at quasi-static speeds smaller than 3 nm/s (Fig. 1h and Fig. S1), which was free from the effect of bulk viscosity [25] and generated a small hysteresis of forces (Fig. 1i and Fig. S1). The resulting symmetric profile confirms that an elastic rather than viscous deformation dominated (Fig. 1f). By contrast, when the jumps were involved, cholesterics recovered to the layered structures simply by retracting the surface from the contact, which showed a self-healing liquid-like property.

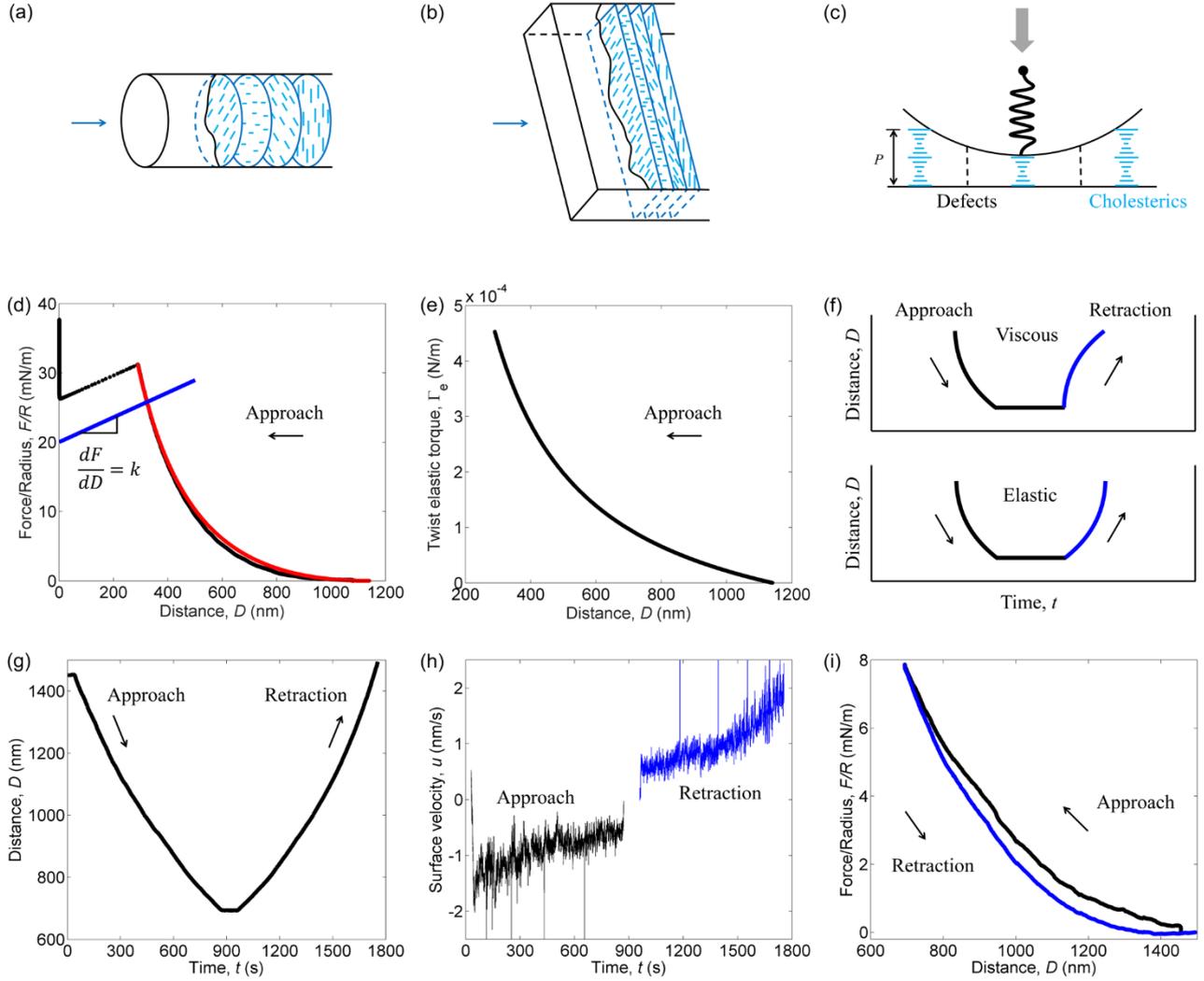

Figure 1 Mechanical winding of cholesterics by surface torque. (a) Schematic diagram of permeative flows in a capillary. (b) Schematic diagram of permeative flows between two plates. (c) Front-view sketch of cholesterics confined in crossed cylinders (Surface Force Balance, SFB). $P = 244$ nm is the pitch of cholesterics. The cylinder radius is $R = 1$ cm. (d) The surface force during the surface approach measured in the SFB. The blue and red lines are the slope line of the spring constant $k = 179$ N/m in (c) and the theoretical plot with Equation S4 (twist elastic constant $K_{22} = 6$ pN), respectively. (e) The twist elastic torque plotted by Equation S7. (f) Distance profiles of viscous and elastic responses. (g) The distance profile during the reversible elastic deformation of cholesterics in the SFB. (h) The surface velocity and (i) force profiles during the surface motion in (g).

When cholesterics are confined in the SFB, the elastic torque is balanced by the surface torque that is analogous to the friction torque in rotational friction. The surface torque is mainly due to the surface anchoring and the surface viscosity [24,26-28]. With strong anchoring, molecules rotate slowly from the easy axis, thus the viscous torque is negligible. Fig. 1e illustrates the corresponding twist elastic torque of the theoretical force profile Fig. 1d. There is a critical surface torque threshold of around $4.5 \times 10^{-4}$ N/m, beyond which the jump occurs. For a certain surface, the critical surface torque is fixed. Still, based on Equation S7, we can obtain a smaller compression ratio of cholesterics by decreasing the twist elastic constant or increasing the pitch.

## 3  Apparent viscosity

The force measurements described above confirmed the elastic nature of cholesterics under confinement with strong anchoring strength. Furthermore, at a quasi-static speed of surface approach (Figs. 1h and 2a, and Fig. S1), the force due to bulk viscosity is negligible [25]. Heretofore, there is no evidence of large viscosity generated by the permeative flows. However, if an effective viscosity is assumed by ignoring the composite details of the measured forces, similar to the Poiseuille flow assumption in permeative flows [5,7-11], the apparent viscosity $\eta_a$ due to the twist elastic deformation can be calculated (see Supplemental Material) as,

$$\eta_a = \frac{1}{6}\gamma_{sa}q_0^2 \frac{D_0^2}{R} = \frac{1}{6}\gamma_{sa}\frac{\Phi_0^2}{R} \tag{1}$$

where $\gamma_{sa}$ is the apparent surface viscosity with the unit of Pa·s·m, $q_0 = \Phi_0/D_0$ is the natural rotational rate of cholesterics, $\Phi_0$ is the total twist angle, $D_0$ is the original surface separation, and $R$ is the radius of the cylinder. This form, i.e., the third type of permeative flows, which is independent of the flow velocity, is very similar to the descriptions of permeative flows in parallel plates and capillaries [5,7-9,11]. Similarly, permeative flows of both cholesterics and smectics in other geometries of channels could be derived.

Fig. 2b shows the apparent viscosity calculated from the experimental forces in Fig. 1d. The data calculated from the experiment scatter because of the scattering velocity (Fig. 2a) but is well matched by the calculation from Equation S10. With more than 70 % strain, the apparent viscosity is 2000 Poise, three orders of magnitude higher than the typical cholesteric bulk viscosity in the order of 1 Poise. If the cholesteric layers were further compressed to 50 nm under infinite anchoring strength, the force $F/R$ would reach 300 mN/m, which would generate an apparent viscosity of $10^5$ P (Fig. 2, c and d). This apparent viscosity is five orders of magnitude higher than the bulk viscosity, which is consistent with the observation in permeative flows. At this point, the maximum surface torque is about 3.4 mN/m (Equation 7), and the anchoring strength is around 2.1 mN/m, with a maximum molecular deviation of π/2 from the easy axis (Equation 6). This anchoring strength is smaller than the maximum theoretical estimation by the dimensional argument [11]. In reality, the anchoring strength on mica surfaces may not be strong enough to sustain such a high compression ratio. However, in the cylindrical capillary, the anchoring torque on the wall accumulates over a long distance, which is possible to generate much larger pinned forces. That is why the ratio between the radius of the capillary and the pitch matters to generate permeative flows [7,18,29]. Furthermore, by calculation, we showed that the apparent viscosity is thickness-dependent under the same strain, exhibiting a parabolic relationship (Fig. 2e). The apparent viscosity is also inversely proportional to the motor speed (Fig. 2f). With the same elastic response, a smaller motor speed achieves a larger apparent viscosity.

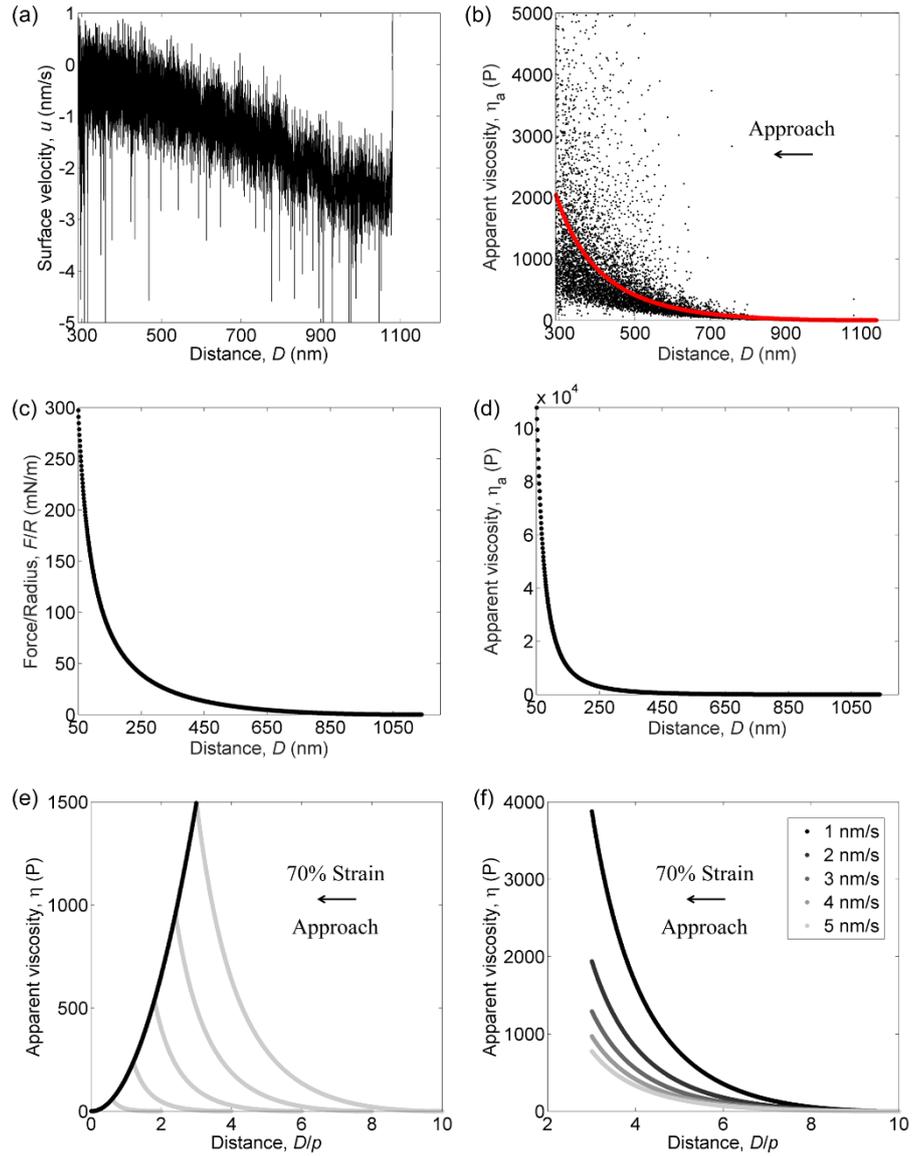

Figure 2 Apparent viscosity of cholesterics. (a) The velocity profile during the approach in Fig. 1d. (b) Apparent viscosity calculated from the force profile in Fig. 1d using Equation S10. The red line represents the theoretical calculation using Equation S10. (c) Theoretical extrapolation of the surface force using Equation S4. (d) Apparent viscosity calculated from the force profile in (c) using Equation S10. (e) Apparent viscosity with various layers calculated by Equation S10. The motor speed of 2.6 nm/s, the same as the experimental value in Fig. 2a, was used. The black line is a parabolic fit of the viscosities at 70% strain. (f) Apparent viscosity with various motor speeds calculated by Equation S10. The apparent viscosity at the same strain is inversely proportional to the motor speed. Symbol $p$ is the half-pitch.

In previous studies, high apparent viscosity accompanied by yield stress [4,12,13] was observed, which was also observed in our experiment (Fig. 1d) with a critical jumping threshold determined by the critical surface torque. Additionally, a study [4] found that when some impurities were present in the samples, small apparent viscosities were measured at low shear rates. This measurement is consistent with our previous results, in which defect nucleation and anchoring strength were affected by water adsorption [24]. In the original paper reporting the high viscosity of cholesterics in the capillary [4], samples with high purity at temperatures above 100 °C were used, and as a result, water and volatile organic molecules were largely avoided.

The rapid increase in surface force and apparent viscosity originates from the twist structure that stores elastic energy, which has been applied to the Paperfuge [30] and artificial muscles [31-34]. More unique mechanical properties are discussed below.

## 4 Strain-stiffening

With strong anchoring, the force per unit area, namely disjoining pressure or stress, and Young's modulus are calculated based on Equation S3 (see Supplemental Material). Fig. 3 shows that both pressure and Young's modulus are not linear as a function of distance, indicating an intrinsic strain-stiffening [35] relationship. The pressure climbs to $2.8 \times 10^4$ Pa with increasing strain to more than 70%. By contrast, Young's modulus under more than 70% strain is approximately $3.8 \times 10^4$ Pa, which is larger than Young's modulus of aqueous foam and granular bed [36], indicating that a large amount of energy can be stored in the twist configuration [31]. Furthermore, both pressure and Young's modulus are thickness-independent, but they get steeper slopes with a thinner thickness (Fig. 3, c and d). This helical structure serves as a liquid spring, which is also vital for mechanical response in the biological system [6].

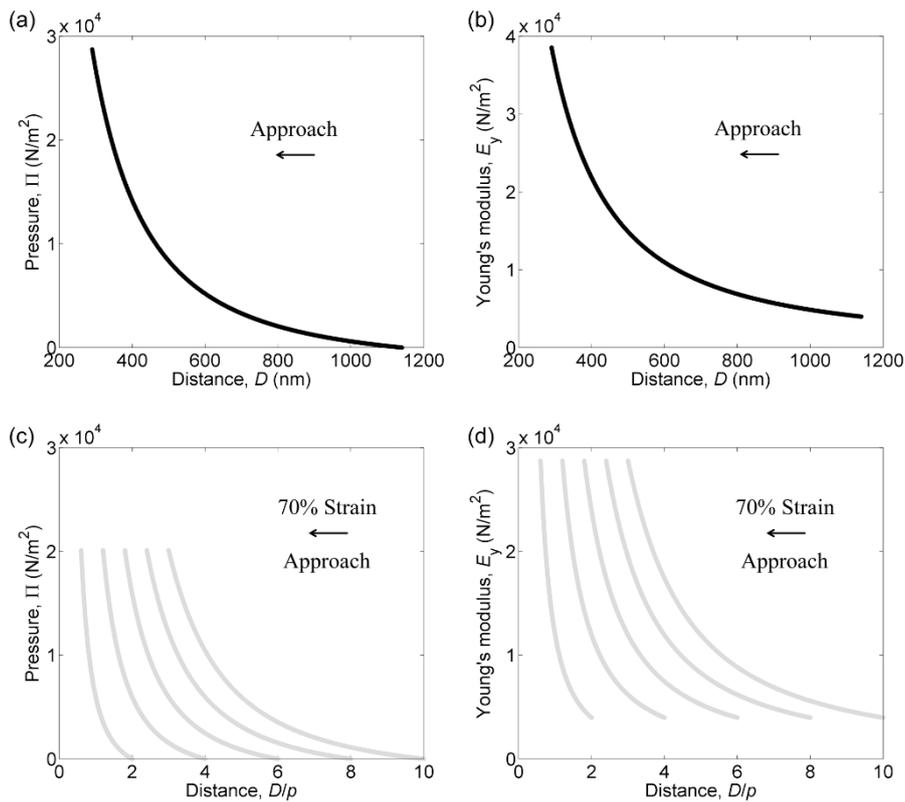

Figure 3 Strain-stiffening of cholesterics. (a) Disjoining pressure profile calculated by Equation S14. (b) Young's modulus profile calculated by Equation S15. Both profiles use the distance range in Fig. 1d. (c) Pressure profiles with various layers up to 70% strain calculated by Equation S14. (d) Young's modulus profiles with various layers up to 70% strain calculated by Equation S15. Symbol $p$ is the half-pitch.

## 5   Nanomotors

The torque obtained in Fig. 1e can store and release power for nanomotors [37]. Here, a cuboid glass rod with dimensions of $a \times a \times l = 5 \times 5 \times 28$ μm$^3$ rather than a cylindrical rod [37] was used for simplification to calculate the generated angular velocity by cholesterics under compression (Fig. 4a). Fig. 4b shows that the maximum torque in Fig. 1e will speed up the angular velocity of the rod to $10^5$ rad/s. This angular velocity is around $10^6$ rpm, which is larger than the speed of the ultracentrifuge and on a par with the theoretical limit by Paperfuge [30]. Additionally, although a thinner thickness obtains a steeper slope for both torque and angular velocity profiles, the torque is thickness-independent and the corresponding angular velocity with the same strain follows a square-root trend (Fig. 4, c and d).

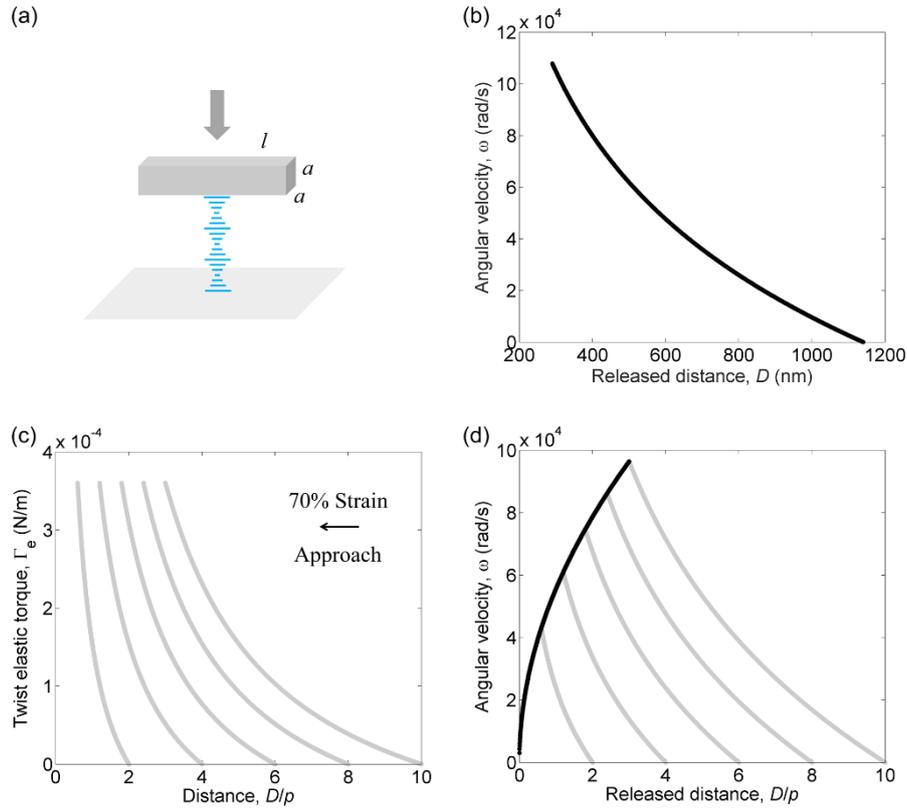

Figure 4 Nanomotors powered by the surface torque. (a) Schematic diagram of a cuboid glass rod compressing cholesterics. The dimension of the rod is $a \times a \times l$. (b) Angular velocity of the rod powered by the surface torque released at various compressing distances, calculated using Equations S18. The initial distance before compression is around 1140 nm, which is the same as the experimental value in Fig. 1e. (c) Torque profiles calculated by Equation S7 with various layers up to 70% strain. (d) Angular velocity of the rod powered by the surface torque profiles in (c). Symbol $p$ is the half-pitch. The black line is the square-root trend line of the angular velocities at 70% strain.

## 6   Discussions

Cholesteric mechanics have been well-developed [38] since the 1950s. For elasticity, the Franck elastic energy governs,

although, at scales comparable to the pitch, the cholesteric structures are too sophisticated to be analyzed [39]. However, two studies [40,41] from the same institute in 2011 still came to opposite conclusions on whether there are oscillatory structural forces in cholesterics. For viscosity, the hydrodynamics in crystals, liquid crystals, and normal fluids have been unified [42]. Nevertheless, the permeative flow has been a puzzle for over 50 years. Coincidentally, the cholesteric-nematic unwinding [11,24,43], permeative flows [5], and Rapini-Papoular anchoring potential [44] were all brought up almost at the same time, and debates about these three concepts have been on-going. It seems that most of the debates point towards the intriguing surface anchoring, namely the boundary conditions. Which anchoring potentials describe the deviation of molecules from the easy axis of the surface is still an open question [45].

Imagine that there is a black box, in which a material is unknown. If we believe that the material is a liquid, we can calculate its effective viscosity with limited measurements of force responses, even though it may be a soft solid. In particular, the visco-elastic duality of liquid crystals makes the feeling of the material more confusing. Learning from the historical lesson on the nature of light, we cannot simply conclude that liquid crystals are fluids. There are instances where liquid crystals behave as pure solids. When it comes to the debates related to molecular constraints, such as surface anchoring and confinement, the length scale really makes a difference. At small scales of $D = 1000$ nm, the Ericksen number [46,47] $Er = \frac{\alpha v D}{K_{22}} = 3 \times 10^{-5}$ is very small, with rotational viscosity of $\alpha = 1$ P, and velocity of $v = 2$ nm/s, indicating the domination of elasticity in cholesterics.

This study may provide a solution to historical puzzles regarding the effect of surface anchoring on cholesteric mechanics. The geometry of crossed cylinders at small surface separation is similar to the geometry with a sphere approaching a flat surface, which amplifies the interacting forces (a factor of *R*/*D*) with tunable determined anchoring conditions. This two-sphere rheology works like the elementary interaction between two crosslinking points in nanocomposite gels [48]. With nano-flakes as additives, it is promising to design flexible yet tough liquid crystal materials without introducing defects [49].

At similar scales in biology, different components that commonly exhibit liquid crystal phases [50] are compartmentalized and confined in functional regions, between which boundaries are ubiquitous. As a result, elasticity also dominates and may adjust the liquid-liquid phase separation [51] in biology. For instance, lipid rafts on the cell membrane [52] may serve as boundaries and play a significant role in the relaxation of membrane tension.

# 7 Conclusions

We have clarified the viscous anomaly in cholesterics with uniform domains confined in the SFB. The cholesteric molecules rotate during the mechanical winding, which is sustained by the surface torque and stores energy in the twist elasticity without dissipating as viscosity. Meanwhile, the enormous viscosity measured in permeative flows is the equivalent conversion of all measured forces, ignoring composite details, into apparent viscosity. The twist structures in cholesterics exhibit an intrinsic strain-stiffening behavior, which underpins the importance of helicities in fundamental mechanical responses in biology. Moreover, the frictional surface torque imposed by the strong anchoring shows promising powers for nanomotors and self-healing artificial muscles.

Our study sheds light on the understanding of the plasticity of liquid crystals [53], anomalous anisotropic diffusion [54,55], the change of the Burgers vector in the Cano wedge [56,57], and viscoelasticity of biological mechanics [1,2].

Meanwhile, some of the research involving viscoelastic materials close to the boundaries may need to be revisited, such as the piezoelectricity along the helical axis of cholesterics [58], and the rheology of defect gels [59].

Acknowledgment

W.Z. is very grateful to S. Perkin who suggested that the measured forces could be due to the equilibrium contribution of cholesterics, and who contributed to the design of experiments and revised the abstract of this work. W.Z. thanks J. Yeomans, R. Lhermerout, J. Hallett, B. Zappone, and R. Bartolino for their helpful discussions. Some contents of this work have been discussed in the Ph.D. thesis titled "Optical and mechanical responses of liquid crystals under confinement (2020)". This paper was supported by the European Research Council (under Grant Nos. ERC-2015-StG-676861 and 674979-NANOTRANS).

*zwhich@outlook.com

# Supplemental Material

## 1 Theory

### 1.1 Free energy

In experiments, the free energy per unit area $G$ mainly consists of twist elastic energy, ignoring the anchoring energy and defect energy under strong anchoring [1],

$$G = \int_0^D \frac{1}{2} K_{22} \left(\frac{\partial \Phi}{\partial D} - q_0\right)^2 dD \tag{1}$$

where $D$ is the surface separation, $K_{22}$ is the twist elastic constant, $\frac{\partial \Phi}{\partial D} = \frac{\Phi}{D}$ is the molecular rotation rate with a total twist angle $\Phi$, and $q_0 = \frac{2\pi}{P}$ is the molecular rotation rate at relaxation. By integrating over the entire distance range, the free energy becomes,

$$G = \frac{1}{2} K_{22} \left(\frac{\Phi}{D} - q_0\right)^2 D \tag{2}$$

With strong anchoring, $\Phi \approx \Phi_0 = q_0 D_0^n$, the twist angle is almost the same as the original one $\Phi_0$ with $n$ layers at the distance $D_0^n$. Thus, the free energy $G^n$ with $n$ layers and the corresponding force $F$ with Derjaguin approximation is calculated as,

$$G^n = \frac{1}{2} K_{22} q_0^2 \frac{(D_0^n - D)^2}{D} \tag{3}$$

$$F = 2\pi R G^n = \pi R K_{22} q_0^2 \frac{(D_0^n - D)^2}{D} \tag{4}$$

where $R$ is the radius of the cylinder.

### 1.2 Surface torque

The twist elastic torque from cholesterics is balanced by the surface torque that consists of anchoring torque and viscous torque.

$$K_{22} \left(\frac{\partial \Phi}{\partial D} - q_0\right) = W \frac{\Phi_0 - \Phi}{2} - \gamma_s \frac{\partial \Phi}{\partial t} \tag{5}$$

where $\gamma_s$ is the surface viscosity with the unit of Pa·s·m, $W$ is the anchoring strength, and $t$ is the time. With strong anchoring, molecules rotate slowly on the surface, and thus the viscous torque is negligible. Therefore, the elastic torque $\Gamma_e$ is balanced by the anchoring torque $\Gamma_a$,

$$\Gamma_e = K_{22} \left(\frac{\Phi}{D} - q_0\right) = \Gamma_a = W \frac{\Phi_0 - \Phi}{2} \tag{6}$$

$$\Gamma_e = K_{22} q_0 \left(\frac{D_0}{D} - 1\right) \tag{7}$$

### 1.3 Apparent viscosity

In the SFB, the total force $F$ is measured by the sum of bulk viscous force and the elastic force [2,3] under the Derjaguin approximation,

$$k(D - D_0 + vt) = F = -\frac{6\pi\eta R^2}{D} \times \frac{dD}{dt} + \pi R K_{22} q_0^2 \frac{(D_0-D)^2}{D} \tag{8}$$

where $k = 179$ N/m is the spring constant, $v$ is the constant motor speed and $\eta$ is the bulk viscosity. With a slow surface approach, the bulk viscous force is negligible [3]. We assume that all the measured forces are equivalent to be responded by an apparent viscosity $\eta_a$, similar to the Poiseuille flow assumption in the calculation of permeative viscosity [4-9]. Therefore, the apparent viscosity is calculated as,

$$-\frac{6\pi\eta_a R^2}{D} \times \frac{dD}{dt} = F \approx \pi R K_{22} q_0^2 \frac{(D_0-D)^2}{D} \tag{9}$$

$$\eta_a \approx -\frac{1}{6} K_{22} q_0^2 \frac{(D_0-D)^2}{R\frac{dD}{dt}} \tag{10}$$

If an apparent surface viscosity $\gamma_{sa}$ is assumed as the boundary condition, Equation S5 is also described by,

$$K_{22}\left(\frac{\partial \Phi}{\partial D} - q_0\right) = -\gamma_{sa} \frac{\partial \Phi}{\partial t} \tag{11}$$

$$K_{22}\left(\frac{\Phi}{D} - q_0\right) = -\gamma_{sa} \frac{\Phi}{D} \cdot \frac{\partial D}{\partial t} \tag{12}$$

The combination of Equations S10 and S12 results in a further simplified apparent viscosity with a strong anchoring limit,

$$\eta_a = \frac{1}{6} \gamma_{sa} q_0^2 \frac{D_0(D_0-D)}{R} \tag{13}$$

At a large compression close to the contact position, the apparent viscosity becomes Equation 1 in the main text.

### 1.4 Strain-stiffening

The disjoining pressure $\Pi$ and Young's modulus $E_y$ are calculated from Equation S3 as,

$$\Pi = -\frac{dG}{dD} = \frac{1}{2} K_{22} q_0^2 \left(\frac{D_0^2}{D^2} - 1\right) \tag{14}$$

$$E_y = \frac{\Pi}{(D_0-D)/D_0} = \frac{1}{2} K_{22} q_0^2 \left(\frac{D_0^2}{D^2} + \frac{D_0}{D}\right) \tag{15}$$

### 1.5 Nanomotors

A cuboid glass rod with dimensions $a \times a \times l = 5 \times 5 \times 28$ μm$^3$ is used to compress cholesterics (Fig. 4a). The moment of inertia $I$ is calculated as,

$$I = \frac{1}{12}m(a^2 + l^2) = \frac{1}{12}\rho a^2 l(a^2 + l^2) \tag{16}$$

where $m$ and $\rho = 2530 \text{ kg/m}^3$ are the mass and density of the glass rod. The rotational kinetic energy is obtained,

$$GA = Gal = \frac{1}{2}I\omega^2 = \frac{1}{24}\rho a^2 l(a^2 + l^2)\omega^2 \tag{17}$$

$$\omega = \sqrt{\frac{24G}{\rho a(a^2+l^2)}} \tag{18}$$

where $A$ is the interaction area of the free energy and $\omega$ is the angular velocity.

## 2 Results

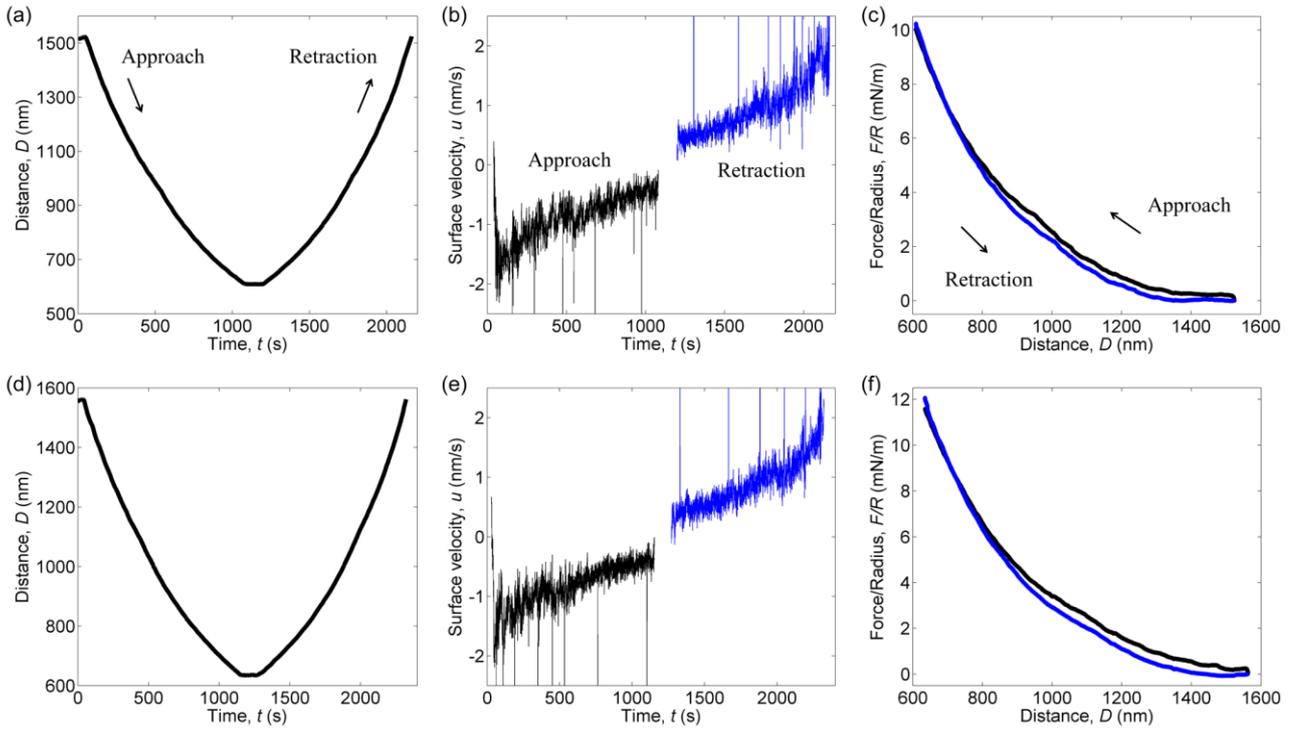

Fig. S1 Reversible deformation of cholesterics. (a) The distance profile. (b) The surface velocity and (c) force profiles during the surface motion in (a). (d-f) Another example of reversible deformation. The spikes in (b) and (d) are due to the data combination from the experimental movies.